\documentclass[prb,twocolumn,superscriptaddress,showpacs,amsmath,amssymb]{revtex4}
\usepackage{amsfonts}
\usepackage{bbm}
\usepackage{graphicx}% Include figure files
\usepackage{dcolumn}% Align table columns on decimal point
\usepackage{bm}% bold math

\begin{document}
\title{Ginzburg-Landau-type theory of non-polarized spin superconductivity }

\author{Peng Lv}
\affiliation{International Center for Quantum Materials, School of Physics, Peking University, Beijing 100871, China}
\author{Zhi-qiang Bao}
\affiliation{University of Texas at Dallas, Department of Physics, Richardson, Texas 75080, USA}
\author{Ai-Min Guo}
\affiliation{Department of Physics, Harbin Institute of Technology, Harbin 150001, China}
\author{X. C. Xie}
\affiliation{International Center for Quantum Materials, School of Physics, Peking University, Beijing 100871, China}
\affiliation{Collaborative Innovation Center of Quantum Matter, Beijing 100871, China}
\author{Qing-Feng Sun}
\email[]{sunqf@pku.edu.cn}
\affiliation{International Center for Quantum Materials, School of Physics, Peking University, Beijing 100871, China}
\affiliation{Collaborative Innovation Center of Quantum Matter, Beijing 100871, China}

\begin{abstract}
Since the concept of spin superconductor was proposed, all the related studies concentrate on spin-polarized case.
Here, we generalize the study to spin-non-polarized case.
The free energy of non-polarized spin superconductor is obtained, and the Ginzburg-Landau-type equations are derived by using the variational method.
These Ginzburg-Landau-type equations can be reduced to the
spin-polarized case when the spin direction is fixed.
Moreover, the expressions of super linear and angular spin currents inside
the superconductor are derived.
We demonstrate that the electric field induced by super spin current
is equal to the one induced by equivalent charge obtained from the second Ginzburg-Landau-type equation, which shows self-consistency of our theory.
By applying these Ginzburg-Landau-type equations,
the effect of electric field on the superconductor is also studied.
These results will help us get a better understanding of the spin
superconductor and the related topics such as Bose-Einstein condensate of magnons and spin superfluidity.
\end{abstract}
\maketitle

\section{Introduction}

Since its original discovery about one century ago,\cite{cplul122-124}
superconductivity has led to an incredible number of surprising phenomena.
Among them, there are several important milestones, such as the London theory,\cite{prs149-71}
the Meissner effect,\cite{Naturwiss21-787} the Ginzburg-Landau (GL) theory,\cite{zetf20-1064} the BCS theory,\cite{pr108-1175} and the Josephson effect.\cite{pl1-251}
Conventional charge superconductivity can be regarded as a superfluid of electric charge,
where electrons form Cooper pairs in the superconductor and condense into the BCS ground
state.\cite{pr108-1175}
Each Cooper pair carries an electric charge $2e$ and is spin singlet for s-wave pairing.
The charge superconductor can support dissipationless charge currents at equilibrium, with the resistance being zero. It is also a perfect diamagnet, which expels external magnetic fields and is known as the Meissner effect.
Electrons have both charge and spin degrees of freedom. The concepts of charge current and voltage can also be applied to the case of spin,
namely spin current and spin voltage. The physical phenomena of the Hall effect\cite{Hall} and the quantum Hall effect\cite{IQHE}
have their counterpart versions, i.e., the spin Hall effect\cite{SHE} and the quantum spin Hall effect.\cite{QSHE1} The spin can thus be considered as the counterpart of the charge, and this correspondence now expands the area of superconductivity.

Recently, the concept of spin superconductivity was proposed.\cite{prb84-214501,prb86-085441,prb87-245427,nc4-2951,epjb86-496,pra91-023633}
The spin superconductivity is a novel quantum state and can be viewed as the counterpart of
the charge superconductivity. The charge superconductor is a superfluid of electric charge as mentioned above,
while the spin superconductor is a superfluid of spin.
The spin superconductor can carry dissipationless flow of spin current, with spin resistance being zero. However, the spin superconductor is a charge insulator and the charge cannot flow through it.
The spin superconductivity may exist in spin-polarized triplet exciton systems of graphene,\cite{prb84-214501,prb86-085441,prb87-245427,nc4-2951,epjb86-496,pra91-023633} $^3$He superfluidity, Bose-Einstein condensate of magnetic atoms, Bose-Einstein condensate of magnons and spinons, and so on.\cite{NP1-111,nature471-83,PRL114-070401}
Exciton is charge neutral and its spin may be either singlet or triplet.
The theory has predicted that the exciton could be realized in a double-layer system,\cite{FNT2-505,JETP}
and later on the experiments have demonstrated the existence of exciton condensate in an electron-hole bilayer system.\cite{PRL68-1196,PRL73-304}
We have also noted that a growing interest has recently arisen in the related topic of spin superfluid,
a superfluid state of spin, which is similar to spin superconducting state.
Theoretical works have proposed several methods to realize dissipationless spin current
in both ferro- and antiferro-magnetic insulators.\cite{prb84-214501,Spin1,Spin2,Spin3}
The influence of long-range dipole interactions on the spin transport properties has been
considered.\cite{long} The phase coherent dc spin transport has shown distinct nonlocal
magnetoresistances and evidence of spin superfluid state.\cite{nonlocal}
More recently, two-fluid theory has been proposed for spin superfluid,\cite{twofluid}
which can be seen as the extension of systems exhibiting U(1) symmetry breaking.
Some theoretical works have also suggested to detect spin superfluid in the $\nu=0$ quantum Hall state of graphene.\cite{prb87-245427,spin-gra}
In the previous works of the spin superconductivity, the London-type theory, the BCS-type theory, the electric Meissner effect, the GL-type theory, and the spin-current Josephson effect have been well established.\cite{prb84-214501,nc4-2951,epjb86-496}
However, these works are referred to spin-polarized case, where the spin direction is fixed by external magnetic fields or magnetic moment of ferromagnetic materials.

The main difference between the spin degree and the charge degree is their dimension.
Spin is a vector in spin space, whereas charge is a scalar in real space.
The spin superconductor can be viewed as Bose-Einstein condensate of charge neutral and spin-nonzero Bosons.
For a system of identical Bosons with spin $f$ ($f$ is a nonzero integer),
the eigenstates of spin operators $\{\mbox{\boldmath $s$}^2,s_{z}\}$ are $|f,m \rangle$ ($m$=$f$,$f-1$,$\cdots$,$-f$).
Each spin state can be characterized by a GL order parameter $\psi_{f,m}$
and the total number of order parameters is $2f+1$.
If the spin direction is completely fixed by external magnetic fields or ferromagnetic moment,
Bosons will only occupy a single spin state $|f,f \rangle$.
All the order parameters vanish except for $\psi_{f,f}$.
Thus, the order parameter is a scalar for the polarized spin superconductor, as discussed in the previous work.\cite{nc4-2951} However, in general, the spin direction is not fixed and can rotate in space.
In this spin-non-polarized case, all the spin states have a probability to be occupied by Bosons.
The order parameter should be a vector $\{\psi_{f,m}\}$
($m$=$f$,$f-1$,$\cdots$,$-f$) in spin space for a non-polarized spin superconductor.

The flow of electric charge generates charge current and similarly the flow of spin produces spin current. The charge only has translational motion, whereas the spin possesses both the translational motion and rotational motion (precession) because it is a vector. As a result, besides a linear spin current which corresponds to the translational motion of the spin, there also exists an angular spin current which describes the rotational motion of the spin.\cite{prb72-245305,prb77-035327}
Regarding the polarized spin superconductor, since the spin direction is fixed by magnetic fields or ferromagnetic moment, there only exists the linear spin current with the angular one being zero. However, when the spin direction is not fixed, both the translational motion and the rotational motion of the spin could exist.

In this paper, we generalize the study to the non-polarized spin superconductivity, where both the linear spin current and the angular one can flow along the system without dissipation. However, the non-polarized spin superconductor is a charge insulator and the charge current cannot flow through the system, because Bose carriers are electric neutral.
The non-polarized spin superconductor may exist in a variety of systems, such as Bose-Einstein condensate of magnetic atoms,\cite{NP1-111,nature471-83,PRL114-070401} $^3$He superfluidity,\cite{add1,add2} and condensate of spinons or magnons in magnetic materials (e.g., graphene with filling factor being $\nu=0$, TlCuCl$_3$, and BaCuSi$_2$O$_6$).\cite{prb84-214501,Spin1,Spin2,Spin3,nonlocal,twofluid,spin-gra,add3,add4,add5}
We write down the free energy and obtain the GL-type equations by using the variational
method. We show that these GL-type equations can be reduced to the
spin-polarized case by fixing the spin direction. Moreover, we obtain the expressions of the super
linear spin current and the super angular one inside the spin superconductor. Besides, we demonstrate that the GL-type equations are self-consistent, by comparing the electric field induced by the super spin current with the one induced by equivalent charge.
We also study the effect of the electric field on the non-polarized spin superconductor by applying the GL-type equations.

The paper is organized as follows.
In Sec. \ref{sec:energy}, we derive the free energy of the
non-polarized spin superconductor.
In Sec. \ref{sec:equation}, we obtain the GL-type equations by minimizing the free energy with the complex conjugate of the order parameter and the electric potential.
In Sec. \ref{sec:polarized}, we show that when ignoring higher-order infinitesimal terms, the spin-non-polarized GL-type equations are naturally reduced to the spin-polarized case in Ref.[\onlinecite{nc4-2951}].
The expressions of the super linear and angular spin currents are derived in Sec. \ref{sec:current}.
In Sec. \ref{sec:field}, we solve the GL-type equations as an application to a concrete example.
Finally, a brief summary is given in Sec. \ref{sec:conclusions}.
The detailed derivation of the GL-type equations is presented in the Appendix.

\section{\label{sec:energy}The free energy}

In the following, we focus on a spin-1 Boson system which corresponds to spin-triplet condensates of spin superconductor and construct the GL-type theory. In this situation,
the order parameters have three components $\psi_{1}({\bf r})$, $\psi_{0}({\bf r})$, and $\psi_{-1}({\bf r})$. For the spin-1 Boson system, the quasi-wave function can be written as:
\begin{small}
\begin{align}\label{eq:2a}
\Psi=\psi_{1}&\left( \begin{array}{c}
1 \\
0 \\
0 \\
\end{array} \right)
+\psi_{0}\left( \begin{array}{c}
0 \\
1 \\
0 \\
\end{array} \right)
+\psi_{-1}\left( \begin{array}{c}
0 \\
0 \\
1 \\
\end{array} \right)
=\left( \begin{array}{c}
\psi_{1} \\
\psi_{0} \\
\psi_{-1} \\
\end{array} \right),
\end{align}
\end{small}
where $\left(\begin{array}{ccc}1, & 0, & 0 \\ \end{array} \right)^T$,
$\left( \begin{array}{ccc} 0, & 1, & 0 \\ \end{array} \right)^T$, and
$\left( \begin{array}{ccc} 0, & 0, & 1 \\ \end{array} \right)^T$
are, respectively, the eigenfunctions of $\sigma_{z}=1$,
$\sigma_{z}=0$, and $\sigma_{z}=-1$ in $\{\mbox{\boldmath $\sigma$}^2,\sigma_{z}\}$
presentation. Note that Eq.(\ref{eq:2a}) can be easily generalized for spin-$N$ system.
We emphasize that although the theory is constructed on the basis of the spin-1 Boson system, the results also hold for all other kinds of non-polarized spin superconductors, e.g., spin-2 system.

In the absence of external electromagnetic fields,
since the free energy of the system is invariant under spin space rotation and gauge transformation,
the free energy of the system can then be constructed.\cite{add6} The free energy is $F_{s}=\int d^3 r f_{s}({\bf r})$, where\cite{add6,add7}
\begin{align}\label{eq:2b}
f_{s}({\bf r})=
&f_{n}+a\psi_{\mu}^{\ast}\psi_{\mu}+\frac{b_{1}}{2}\psi_{\mu}^{\ast}\psi_{\mu}^{\ast}\psi_{\nu}\psi_{\nu}+
\frac{b_{2}}{2}\psi_{\mu}^{\ast}\psi_{\nu}^{\ast}\psi_{\mu}\psi_{\nu} \nonumber \\
&+\frac{1}{2m}(i\hbar\nabla_{j})\psi_{\mu}^{\ast}(-i\hbar\nabla_{j})\psi_{\mu}.
\end{align}
Here, $f_s$ ($f_{n}$) is the density
of the free energy of the spin superconducting (normal) state. $a$, $b_{1}$, and $b_{2}$ are the parameters in the GL-type
theory, which also appear in the charge superconductor. Since the
order parameter $\Psi$ is a vector, we need to contract it with its dual partner, i.e., $\Psi^\ast$. $a\psi_{\mu}^{\ast}\psi_{\mu}$ is the contraction between two wave functions, while $\frac{b_{1}}{2}\psi_{\mu}^{\ast}\psi_{\mu}^{\ast}\psi_{\nu}\psi_{\nu}$
and $\frac{b_{2}}{2}\psi_{\mu}^{\ast}\psi_{\nu}^{\ast}\psi_{\mu}\psi_{\nu}$
are the contractions among four wave functions. Here, each quartic term corresponds to one different way by contracting the subscript indices. Notice that similar contractions have been used in the free energy of the $^3$He superfluidity.\cite{He,Hebook}
In fact, the third and fourth terms in Eq.(\ref{eq:2b}) are the same as Refs.[\onlinecite{add6,add7}].
The last term of $\frac{1}{2m}(i\hbar\nabla_{j})\psi_{\mu}^{\ast}(-i\hbar\nabla_{j})\psi_{\mu}$
can be regarded as the kinetic energy of the system. Analogous contractions can also be found in the free energy of the spin-polarized superconductor.\cite{nc4-2951} The difference is that the wave function is changed from
single component to three components. We stress that $\mu$ in
$\psi_{\mu}$ is an index in spin space, whereas $j$ in $\nabla_{j}$
(component of $\nabla$) is an index in real space. The index contraction can only be performed in the same space, i.e., the index in spin space cannot be contracted with the index in real space.

In the presence of a magnetic field ${\mathbf B}$ and an electric field ${\mathbf E}$, the density of the free energy becomes:
\begin{align}\label{eq:2c}
f_{s}=&f_{n}+a\psi_{\mu}^{\ast}\psi_{\mu}+\frac{b_{1}}{2}\psi_{\mu}^{\ast}\psi_{\mu}^{\ast}\psi_{\nu}\psi_{\nu}+
\frac{b_{2}}{2}\psi_{\mu}^{\ast}\psi_{\nu}^{\ast}\psi_{\mu}\psi_{\nu} \nonumber \\
&+\frac{[(-i\hbar\nabla-\alpha_{0}\mbox{\boldmath $\sigma$}\times{\mathbf E})\Psi]^{\dagger}\cdot(-i\hbar\nabla-\alpha_{0}\mbox{\boldmath $\sigma$}\times{\mathbf E})\Psi}{2m} \nonumber \\
&+\Psi^{\dagger}g\mu_{B}\mbox{\boldmath $\sigma$}\cdot{\mathbf B}\Psi+\frac{1}{2}\epsilon_{0}{\mathbf E}^2+\frac{{\mathbf B}^2}{2\mu_{0}}.
\end{align}
Here, $\Psi^{\dagger}g\mu_{B}\mbox{\boldmath $\sigma$}\cdot{\mathbf B}\Psi$ is the Zeeman energy, $g$ is the Lande factor, $\mu_{B}$ is the Bohr magneton, and $\mbox{\boldmath $\sigma$}$ is the Pauli matrices in spin-1 representation (not in spin-1/2 representation).  $\frac{1}{2}\epsilon_{0}{\mathbf E}^2$ and $\frac{{\mathbf B}^2}{2\mu_{0}}$
are the energies of the electric field and the magnetic field, respectively.
For the kinetic energy, the term $-i\hbar\nabla$ is replaced by
$-i\hbar\nabla-\alpha_{0}\mbox{\boldmath $\sigma$}\times{\mathbf E}$, because the momentum will be changed from ${\mathbf p}$ to
${\mathbf p}-\alpha_{0}\mbox{\boldmath $\sigma$}\times{\mathbf E}$
when the magnetic moment lies in an electric field, with $\alpha_{0}$ the spin-orbit coupling strength.\cite{nc4-2951,prb95-187203}
In fact, when we consider the Dirac equation and take the non-relativistic limit, the canonical kinetic momentum is written as
$\mathbf{P}_{canonical}=-i\hbar\nabla+\frac{e}{c}{\mathbf A}-\alpha_{0}\mbox{\boldmath $\sigma$}\times{\mathbf E}$. For a charge neutral carrier with nonzero spin, the kinetic energy is just the one mentioned above. While for a charge carrier with zero spin, the kinetic energy is reduced to the case discussed by Ginzburg and Landau.\cite{zetf20-1064}
Note that the free energy in Eqs.(\ref{eq:2b}) and (\ref{eq:2c}) can also describe pseudo-spin-orbit-coupled Bose-Einstein condensates, where the pseudo-spin-orbit coupling can be realized by Raman coupling of atomic hyperfine states and ${\mathbf E}$ is not real electric field as well.\cite{nature471-83,PRL114-070401,add8}
However, in the present work, we consider real spin-orbit coupling which originates from moving of a magnetic moment in an external electric field.

\section{\label{sec:equation}The GL-type equations}

In this section, we use the variational method to obtain the GL-type equations.
By minimizing the free energy $F_{s}$ with respect to the complex conjugate of the order parameter $\Psi^{\dagger}$, we can get the first GL-type equation (see the Appendix):
\begin{align}\label{eq:3a}
&a\Psi+b_{1}\Psi^{\ast}(\Psi^{T}\Psi)+b_{2}\Psi(\Psi^{\dagger}\Psi) \nonumber \\
&+\frac{1}{2m}(-i\hbar\nabla-\alpha_{0}\mbox{\boldmath $\sigma$}\times{\mathbf E})^2\Psi+g\mu_{B}\mbox{\boldmath $\sigma$}\cdot{\mathbf B}\Psi=0.
\end{align}

This first GL-type equation is a nonlinear equation which describes the order parameter $\Psi$. Provided that both the electric field $\mathbf E$ and the magnetic field $\mathbf B$ are given, the order parameter $\Psi$ can be obtained inside the spin superconductor by solving the first GL-type equation. Therefore, one can investigate the influence of electromagnetic field on the order parameter $\Psi$ by considering the first GL-type equation. Notice that $\Psi$ is a vector rather than a scalar in spin space. This indicates that the angular spin current should be finite and new physical phenomena may appear as discussed below.

When we minimize the free energy $F_{s}$ with respect to the electric potential $\varphi$, the second
GL-type equation can be obtained (see the Appendix):
\begin{align}\label{eq:3b}
\rho=-\nabla\cdot\left\{\left[\frac{{\mathbf p}-\alpha_{0}\mbox{\boldmath $\sigma$}\times{\mathbf E}
}{2m}\Psi\right]^{\dagger}\times\alpha_{0}\mbox{\boldmath $\sigma$}\Psi\right\}+c.c..
\end{align}
Here $\rho$ is the equivalent charge density.

This second GL-type equation provides the electromagnetic response of the spin superconductor. After obtaining the order parameter $\Psi$ from the first GL-type equation, the equivalent charge distribution inside the spin superconductor can be obtained by substituting $\Psi$ into the second GL-type equation. Then, one can calculate the electric field induced by the super spin current.
It is now clear that the second GL-type equation describes the equivalent charge generated by the super spin current. This is completely different from the second GL equation of the charge superconductor, which describes the super charge current itself.\cite{zetf20-1064}

Eqs.(\ref{eq:3a}) and (\ref{eq:3b}) are the main results of this paper. We point out that these two equations are universal for all kinds of the non-polarized spin superconductors, although they are derived from the spin-1 Boson system. These GL-type equations are very important for studying the physical properties of the spin superconductors, including the spin-current Josephson effect, the proximity effect, and the electromagnetic response.
They can be used to describe different spin-superconductor systems, ranging from spin-non-polarized system to partial spin-polarized one and completely spin-polarized one.
In the following section, the electromagnetic response of a non-polarized spin superconductor is investigated by using these GL-type equations.

\section{\label{sec:polarized}Spin-polarized limit}

Before we proceed, it is necessary to demonstrate that the above GL-type equations can be reduced to the spin-polarized case by fixing the spin direction.\cite{nc4-2951} Provided that the spin direction is fixed by an external magnetic field ${\mathbf B}=B{\mathbf e}_{z}$ in the$\ \rm z$ direction, the bosons will occupy a single state $|1,1 \rangle$. In this situation, all the components of the order parameter $\Psi$ vanish except for $\psi_{1}$. Then, the quasi-wave function can be expressed as:
\begin{align}\label{eq:4a}
\Psi=\left( \begin{array}{ccc}
\psi_{1} ,& 0,
0 \\
\end{array} \right)^T.
\end{align}

By substituting Eq.(\ref{eq:4a}) into the first GL-type equation, i.e., Eq.(\ref{eq:3a}), we derive the first component:
\begin{align}\label{eq:4b}
[a+(b_{1}+b_{2})|\psi_{1}|^2]\psi_{1}+\frac{1}{2m}(-i\hbar\frac{\partial}{\partial x}-\alpha_{0}\partial_{y}\varphi)^2\psi_{1}\nonumber \\
+\frac{1}{2m}(-i\hbar\frac{\partial}{\partial y}+\alpha_{0}\partial_{x}\varphi)^2\psi_{1}+\frac{1}{2m}(-i\hbar\frac{\partial}{\partial z})^2\psi_{1}\nonumber \\
+\frac{\alpha_{0}^2}{4m}[2(\partial_{z}\varphi)^2+(\partial_{x}\varphi)^2+(\partial_{y}\varphi)^2]\psi_{1}=0.
\end{align}
By replacing $b_{1}+b_{2}$ with $b$ and neglecting the last term of $\alpha_{0}^2$, Eq.(\ref{eq:4b}) can be written as:
\begin{align}\label{eq:4c}
[a+b|\psi|^2]\psi+\frac{1}{2m}(-i\hbar\frac{\partial}{\partial x}-\alpha_{0}\partial_{y}\varphi)^2\psi\nonumber \\
+\frac{1}{2m}(-i\hbar\frac{\partial}{\partial y}+\alpha_{0}\partial_{x}\varphi)^2\psi+\frac{1}{2m}(-i\hbar\frac{\partial}{\partial z})^2\psi=0.
\end{align}
This is just the equation in the spin-polarized case.\cite{nc4-2951} Let us estimate the magnitude of the last term of $\alpha_{0}^2$ and compare it with the momentum. The magnitude of the momentum $\hbar k$ is approximately $\hbar/\delta r$, with $\hbar$ the reduced Plank constant and $\delta r$ the characteristic length of the order parameter $\psi_{1}$. $\alpha_{0}E=e\hbar E/2mc^2$ is the contribution of the spin-orbit coupling term, where $e$ is the elementary charge, $E$ is the magnitude of the electric field, $m$ is the mass of a boson, and $c$ is the speed of light. By choosing the typical values of $\delta r=10^3\ \rm nm$, $E=10^6\ \rm V/m$, and $m=m_{e}$, $\alpha_{0}E/\hbar k$ is approximately $10^{-6}$, which is much smaller than 1.

Additionally, by substituting Eq.(\ref{eq:4a}) into the second GL-type equation, i.e., Eq.(\ref{eq:3b}), one obtains:
\begin{eqnarray}\label{eq:4d}
\rho&= \frac{1}{2m}\nabla\cdot & \left(-i\hbar\alpha_{0}(\psi_{1}\partial_{y}\psi_{1}^{\ast}-\psi_{1}^{\ast}\partial_{y}\psi_{1})-
3\alpha_{0}^2(\partial_{x}\varphi)|\psi_{1}|^2, \right.\nonumber  \\
& & i\hbar\alpha_{0}(\psi_{1}\partial_{x}\psi_{1}^{\ast}-\psi_{1}^{\ast}\partial_{x}\psi_{1})-
3\alpha_{0}^2(\partial_{y}\varphi)|\psi_{1}|^2,  \nonumber  \\
& & \left. -2\alpha_{0}^2(\partial_{z}\varphi)|\psi_{1}|^2\right),
\end{eqnarray}
which is the same as the spin-polarized case.\cite{nc4-2951}

\section{\label{sec:current}The super linear spin current and super angular spin current}

In this section, we derive the relation between the super spin current and the quasi-wave function $\Psi({\bf r})$.
For the polarized spin superconductor, the linear spin current is a vector, whereas the angular one is zero. However, it is very different for the spin-non-polarized case, because the spin direction can rotate. In this situation, both the super linear spin current and the super angular one can exist in general. The linear spin current is a tensor, whereas the angular one is a vector.\cite{prb72-245305}
Firstly, we linearize the first GL-type equation (\ref{eq:3a}):\cite{book1,book2,book3,nc4-2951}
\begin{align}\label{eq:5a}
a\Psi+\frac{1}{2m}(-i\hbar\nabla-\alpha_{0}\mbox{\boldmath $\sigma$}\times{\mathbf E})^2\Psi+g\mu_{B}\mbox{\boldmath$\sigma$}\cdot{\mathbf B}\Psi=0.
\end{align}
Eq.(\ref{eq:5a}) can be rewritten as:
\begin{align}\label{eq:5b}
\Big[\frac{1}{2m}(-i\hbar\nabla-\alpha_{0}\mbox{\boldmath $\sigma$}\times{\mathbf E})^2+g\mu_{B}\mbox{\boldmath $\sigma$}\cdot{\mathbf B}\Big]\Psi=-a\Psi.
\end{align}
By defining a Hamiltonian $H$ as:
\begin{align}\label{eq:5c}
H=\frac{1}{2m}(-i\hbar\nabla-\alpha_{0}\mbox{\boldmath $\sigma$}\times{\mathbf E})^2+g\mu_{B}\mbox{\boldmath $\sigma$}\cdot{\mathbf B},
\end{align}
it is evident that Eq.(\ref{eq:5b}) can be expressed as $H\Psi=-a\Psi$, which is analogous to a Schr\"{o}dinger equation. As a result, $-a$ is the eigenvalue of the Hamiltonian $H$ and all the information of the system can be obtained by solving this Schr\"{o}dinger-like equation. By using the Heisenberg equation of motion, we get:
\begin{align}\label{eq:5d}
{\mathbf v}=&\frac{d}{dt}{\mathbf r}=\frac{1}{i\hbar}\big[{\mathbf r}, H\big] \nonumber \\
=&\frac{1}{i\hbar}\big[{\mathbf r}, \frac{1}{2m}({\mathbf p}-\alpha_{0}\mbox{\boldmath $\sigma$}\times{\mathbf E})^2+g\mu_{B}\mbox{\boldmath $\sigma$}\cdot{\mathbf B}\big] \nonumber \\
=&\frac{1}{m}({\mathbf p}-\alpha_{0}\mbox{\boldmath $\sigma$}\times{\mathbf E}).
\end{align}
By employing ${\mathbf J}_{s}={\rm Re}[\Psi^{\dagger}{\mathbf v}{\mathbf s}\Psi]$,\cite{prb72-245305,Science301-1348,prb68-241315,prl92-126603}
the linear spin current can be written as:
\begin{align}\label{eq:5e}
{\mathbf J}_{s}={\rm Re}[{\Psi^{\dagger}\frac{1}{m}({\mathbf p}-\alpha_{0}\mbox{\boldmath $\sigma$}\times{\mathbf E})\mathbf s}\Psi].
\end{align}
Here the spin ${\mathbf s}=\hbar \mbox{\boldmath $\sigma$}$.
We emphasize that the results are the same even if the GL-type equation is not linearized, because the non-linear terms (the second and the third terms) in Eq.(\ref{eq:3a}) do not contain the momentum operator ${\mathbf p}$.

As we know that in the presence of a magnetic field ${\mathbf B}$, the energy of a magnetic moment ${\mathbf m}$ is $-{\mathbf m}\cdot{\mathbf B}$ and the Larmor precession will occur, with the frequency proportional to the magnitude of the magnetic field. The Hamiltonian in Eq.(\ref{eq:5c}) can be rewritten as:
\begin{align}\label{eq:5f}
H=&\frac{1}{2m}({\mathbf p}-\alpha_{0}\mbox{\boldmath $\sigma$}\times{\mathbf E})^2+g\mu_{B}\mbox{\boldmath $\sigma$}\cdot{\mathbf B} \nonumber \\
\approx&\frac{{\mathbf p}^2}{2m}+\mbox{\boldmath $\sigma$}\cdot\big[\frac{\alpha_{0}}{m}({\mathbf p}\times{\mathbf E})+g\mu_{B}{\mathbf B}\big].
\end{align}
Here, the quadratic term of $\alpha_0$ is ignored, as it is higher-order infinitesimal. As compared with the Larmor precession, we know that the spin could rotate and the frequency operator ${\bf \omega}$ is $\frac{1}{\hbar}\big[\frac{\alpha_{0}}{m}({\mathbf p}\times{\mathbf E})+g\mu_{B}{\mathbf B}\big]$.
Then, the super angular spin current is defined as:\cite{prb72-245305}
\begin{align}\label{eq:5g}
{\mathbf J}_{\omega}={\rm Re}\Psi^{\dagger}\{\frac{1}{\hbar}\big[\frac{\alpha_{0}}{m}({\mathbf p}\times{\mathbf E})+g\mu_{B}{\mathbf B}\big]\times\mbox{\boldmath $s$}\}\Psi.
\end{align}

Here both the expressions of the super linear spin current Eq.(\ref{eq:5e}) and the super angular one Eq.(\ref{eq:5g}) are consistent with Ref.[\onlinecite{prb72-245305}]. Besides, one can also confirm that ${\mathbf J}_{s}$ and ${\mathbf J}_{\omega}$
satisfy the quantum spin continuity equation $\frac{d}{dt} {\mathbf s}({\bf r},t)=-\nabla \cdot {\mathbf J}_{s} + {\mathbf J}_{\omega}$,\cite{prb72-245305} where ${\mathbf s}({\bf r},t)= \Psi^{\dagger}({\bf r},t) {\mathbf s}\Psi({\bf r},t)$ is the local spin density at position ${\bf r}$ and time $t$. For the steady state, we have $\frac{d}{dt} {\mathbf s}({\bf r},t) =0$.
This spin continuity equation arises from the invariance of the spin magnitude $|{\mathbf s}|$, i.e., the spin magnitude of a boson remains constant with $|{\mathbf s}|=\hbar$ no matter how the boson is translated or rotated.\cite{prb72-245305}

\begin{figure}[h]
\includegraphics[width=0.7\columnwidth]{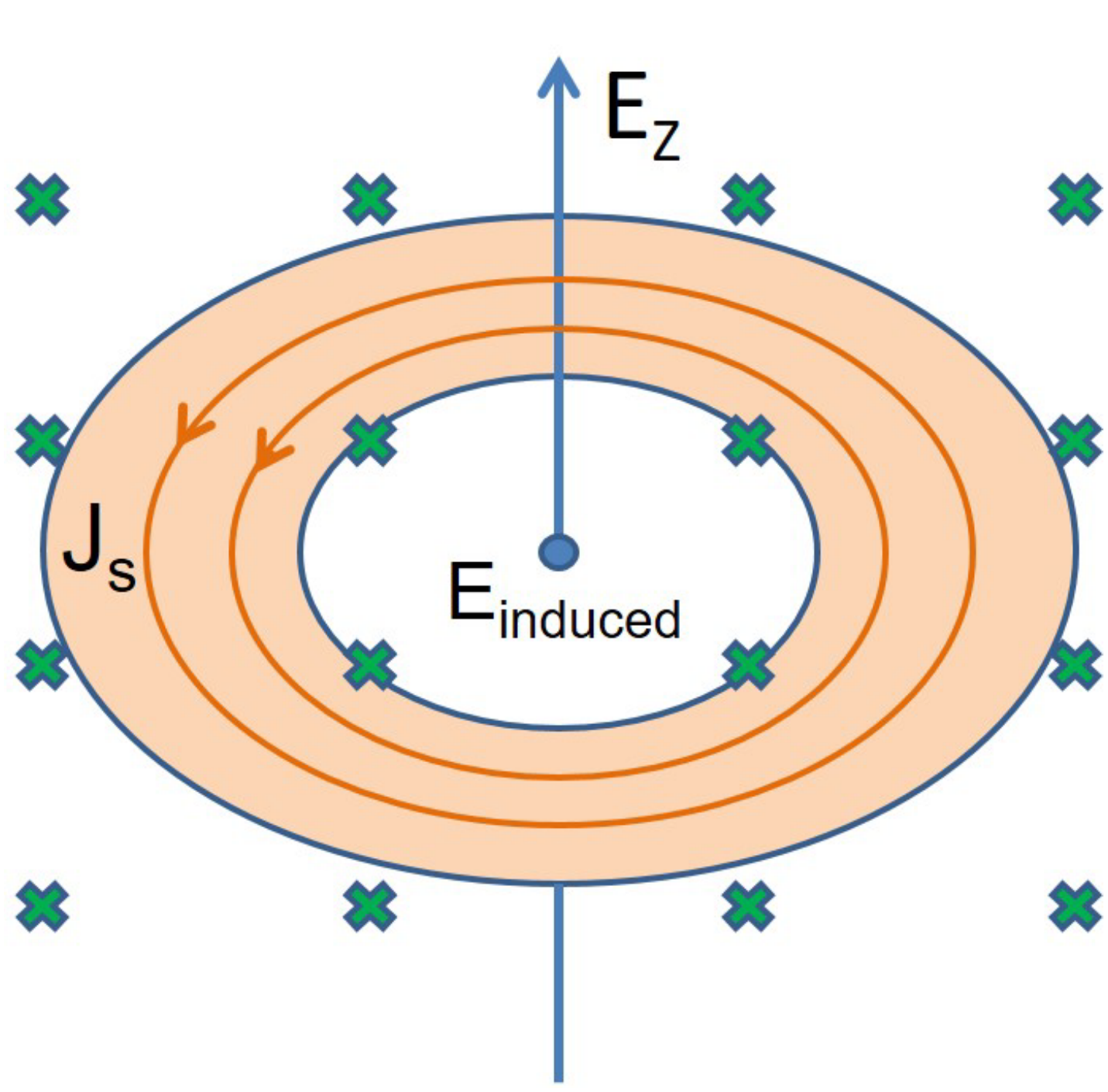}
\caption{(Color online) Schematic diagram of an annulus device of the spin superconductor. A uniform electric field ${\mathbf E}$ is perpendicular to the surface of the annulus, which is very thin and can be regarded as a quasi-2D system. The green crosses represent the electric field ${\mathbf E}$ and the orange circles with arrows denote the super spin current. The super spin current can produce an additional electric field. Figs.4 and 5 show this additional electric field along the center line at ${\bf r}=0$, which is perpendicular to the annulus, and along the radial direction at $z=0$, respectively.} \label{fig1}
\end{figure}

\section{\label{sec:field} the application of the GL equations}

In the above sections, we have constructed the GL-type theory and obtained the expressions of the super spin current inside the non-polarized spin superconductor. As an example, next the GL-type equations are solved in a simple system. We consider an annulus setup with inner radius $R_{0}$ and outer radius $R_{1}$, where the thickness $d$ of the annulus is very small. As a result, this annulus can be considered as a quasi-2D system. When the annulus possesses the spin superconducting phase, the super spin current can be generated in the presence of a constant electric field ${\mathbf E}=E_{0}{\mathbf e}_{z}$ which is perpendicular to the surface of the annulus, as illustrated in Fig.\ref{fig1}. Here, we consider the situation that the system enters into the superconducting phase but is near the critical point. In this case, the superfluid density, i.e., $\Psi^{\dagger} \Psi$, is very small. Then, the non-linear terms can be neglected in the first GL-type equation (\ref{eq:3a}) and one obtains:
\begin{align}\label{eq:6a}
\frac{1}{2m}(-i\hbar\nabla-\alpha_{0}\mbox{\boldmath $\sigma$}\times{\mathbf E})^2\Psi=-a\Psi.
\end{align}
Since the annulus has cylindrical symmetry, it is convenient to study the electromagnetic response under the cylindrical coordinate system. Then, the differential operator is $\nabla = (\frac{\partial}{\partial r},
\frac{1}{r}\frac{\partial}{\partial \theta}, \frac{\partial}{\partial z} )$
and the Pauli matrices in spin-1 representation are expressed as:
\begin{align}\label{eq:6b}
&\sigma_{r}=\frac{1}{\sqrt{2}}\left( \begin{array}{ccc}
0 & e^{-i\theta} & 0 \\
e^{i\theta} & 0 & e^{-i\theta} \\
0 & e^{i\theta} & 0 \\
\end{array} \right), \nonumber \\
&\sigma_{\theta}=\frac{1}{\sqrt{2}}\left( \begin{array}{ccc}
0 & -ie^{-i\theta} & 0 \\
ie^{i\theta} & 0 & -ie^{-i\theta} \\
0 & ie^{i\theta} & 0 \\
\end{array} \right), \nonumber \\
&\sigma_{z}=\left( \begin{array}{ccc}
1 & 0 & 0 \\
0 & 0 & 0 \\
0 & 0 & -1 \\
\end{array} \right).
\end{align}
It should be noted that under the cylindrical coordinate system, the wave function is different from that in Eq.(\ref{eq:2a}). In this situation, the wave function can be easily obtained by performing a unitary transformation on Eq.(\ref{eq:2a}) with the unitary matrix $U=diag(e^{-i\theta}, 1, e^{i\theta})$. This unitary matrix can transform the wave function from the Cartesian coordinate system to the cylindrical one. Then, the wave function in the cylindrical coordinate system is written as:
\begin{align}\label{eq:6c}
\Psi=U\left( \begin{array}{c}
\psi_{1} \\
\psi_{0} \\
\psi_{-1} \\
\end{array} \right)=\left( \begin{array}{c}
\psi_{1}e^{-i\theta} \\
\psi_{0} \\
\psi_{-1}e^{i\theta} \\
\end{array} \right).
\end{align}
By substituting Eqs.(\ref{eq:6b}) and (\ref{eq:6c}) into Eq.(\ref{eq:6a}), the GL quasi-wave function $\Psi$ can be numerically calculated, where $\Psi$ is chosen to be the ground state. Notice that since $|\Psi|^2$ corresponds to the superfluid density $n_{s}$ of the system, the normalization of the quasi-wave function $\Psi$ is $\int d^3r |\Psi|^2=N$, with $N$ the number of condensates of the whole system. In the numerical calculations presented below, we set the inner radius $R_{0}=0.3\ \rm mm$, the outer radius $R_{1}=0.6\ \rm mm$, and the thickness $d=0.03\ \rm mm$. We take the strength of the electric field $E_{0}=-10^{6}\ \rm V/m$, the effective mass $m=0.1m_{e}$, and the superfluid density $n_{s}=10^{28}\ \rm /m^3$.

\begin{figure}
\includegraphics[width=1.0\columnwidth]{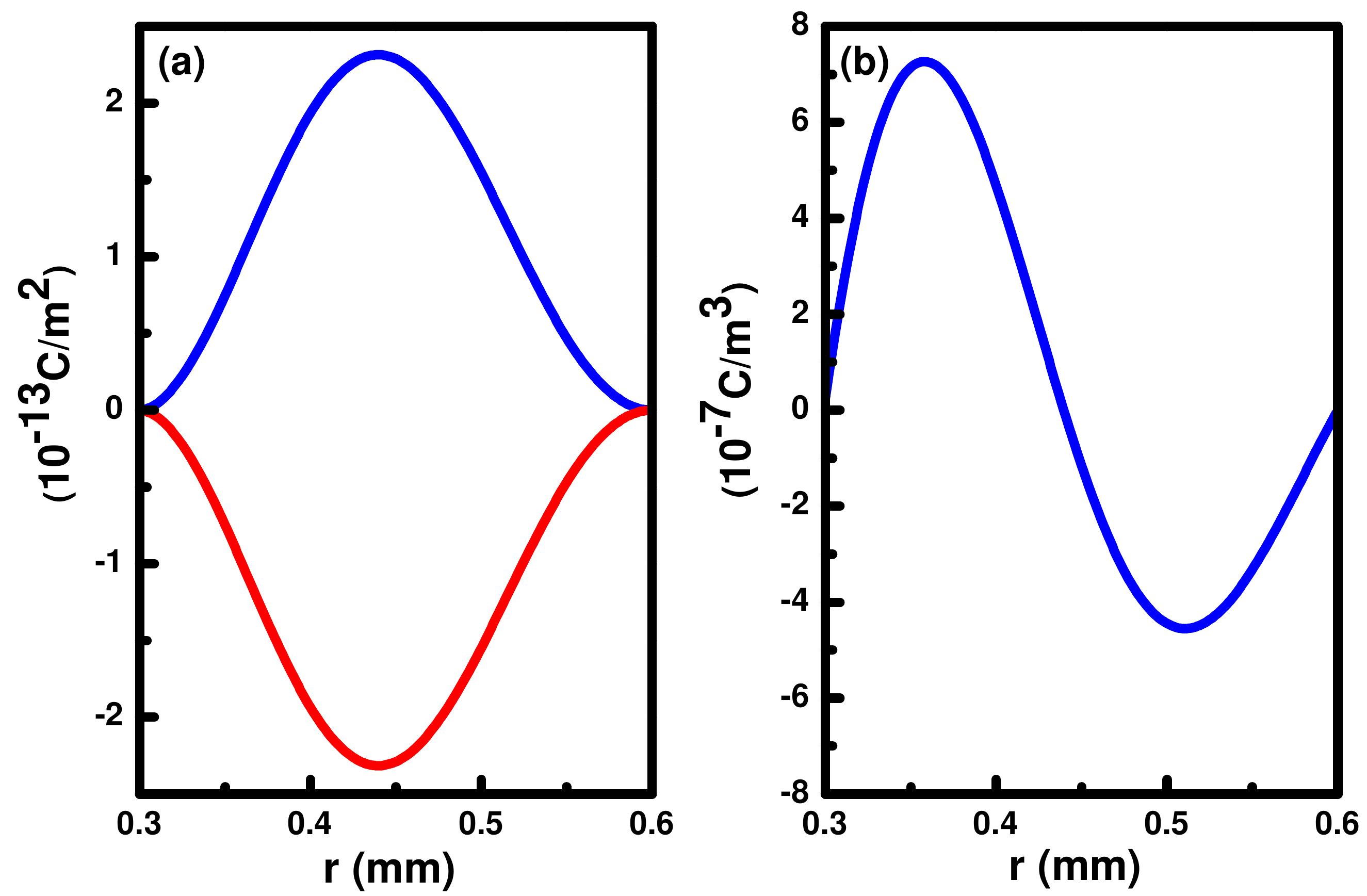}
\caption{(Color online) (a) Equivalent charge density in the upper surface with $z= d/2$ (red line) and in the lower one with $z=- d/2$ (blue line) of the annulus. (b) Equivalent bulk charge density inside the annulus with $z= 0$. The results are calculated at different sites along the radial direction.} \label{fig2}
\end{figure}

After obtaining the quasi-wave function $\Psi$, the equivalent charge distribution inside the annulus can be calculated by substituting $\Psi$ into Eq.(\ref{eq:3b}). Figs.\ref{fig2}(a) and 2(b) display, respectively, the distribution of the equivalent surface charge density $\sigma$ and the equivalent bulk charge density $\rho$ along the radial direction. The magnitude of the equivalent charge density is about $10^{-13} \ \rm C/m^2$ and $10^{-7}\ \rm C/m^3$ in the surface and bulk of the annulus, respectively. The sum of all the induced equivalent charge is zero, because of the charge conservation.

\begin{figure}[h]
\includegraphics[width=1.0\columnwidth]{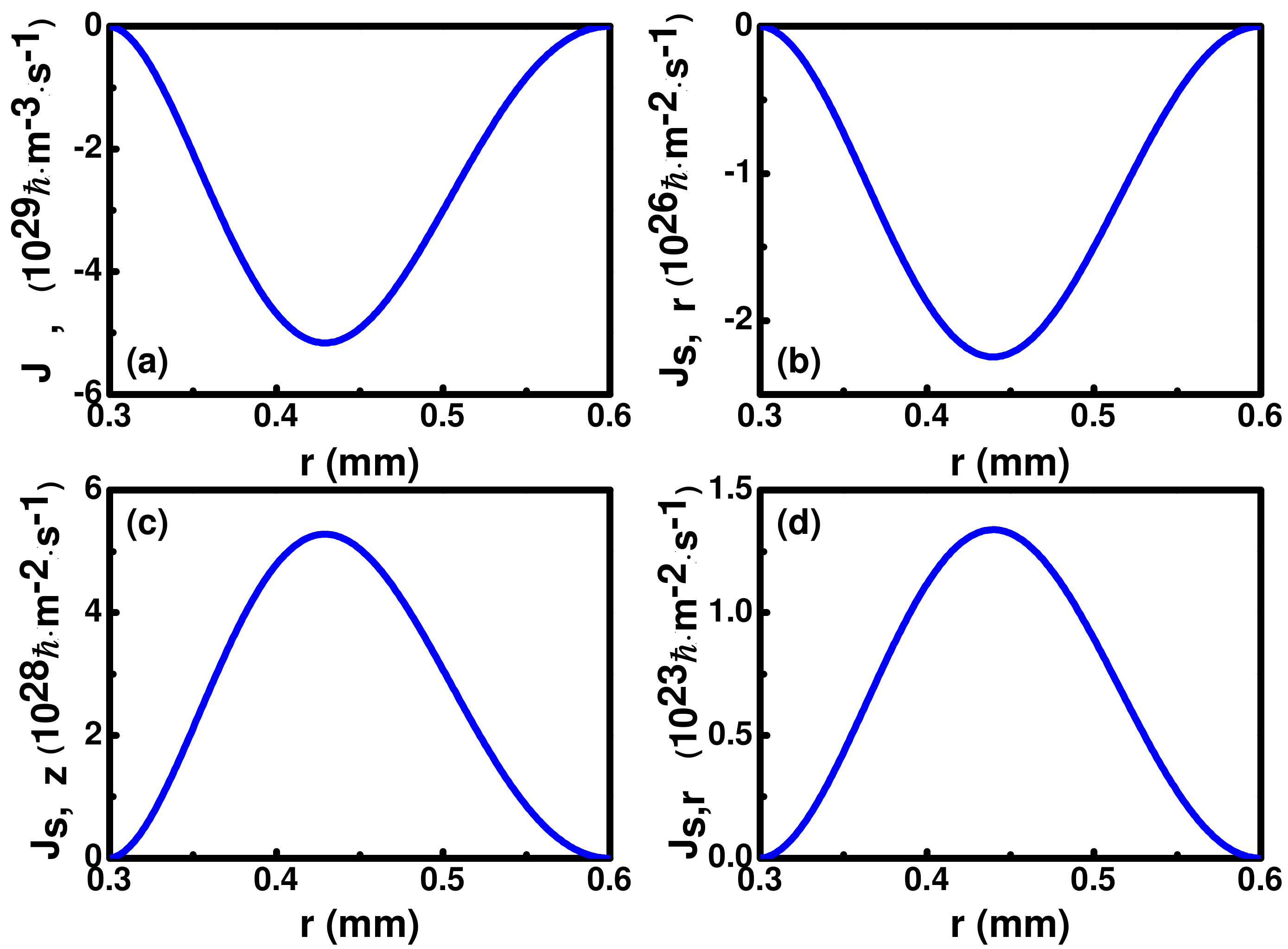}
\caption{(Color online) Super spin current density distribution along the radial direction in the middle of the annulus ($z=0$).
(a)-(d) denote ${\mathbf J}_{\omega,\theta}$, ${\mathbf J}_{{s},\theta r}$, ${\mathbf J}_{{s},\theta z}$, and ${\mathbf J}_{{s},r \theta}$, respectively.} \label{fig3}
\end{figure}

Besides, the super spin current distribution inside the annulus can be calculated by substituting $\Psi$ into Eqs.(\ref{eq:5e}) and (\ref{eq:5g}). Fig.\ref{fig3} shows the super spin current density distribution along the radial direction inside the annulus. As mentioned above, the angular spin current is a vector and the linear one is a tensor. As a result, the spin current has twelve components. However, only four components, including ${\mathbf J}_{\omega,\theta}$, ${\mathbf J}_{{s},\theta r}$, ${\mathbf J}_{{s},\theta z}$, and ${\mathbf J}_{{s},r \theta}$, are nonzero, while all the other components are zero, owing to the cylindrical symmetry of the annulus. Here, for example, ${\mathbf J}_{{s},\theta r}$ represents the spin-1 boson which moves along the angular direction with its spin direction parallel to the radial direction.\cite{prb72-245305} Although the magnitudes of these four components are different from each other, all of them are zero at the interfaces of $r=R_0$ and $R_1$, and are large near the middle as expected. The corresponding superfluid velocity varies from $10^{-5}\ \rm m/s$ to $5\ \rm m/s$ for different components, with the rotational frequency about 50$\ \rm Hz$.

\begin{figure}[h]
\includegraphics[width=0.8\columnwidth]{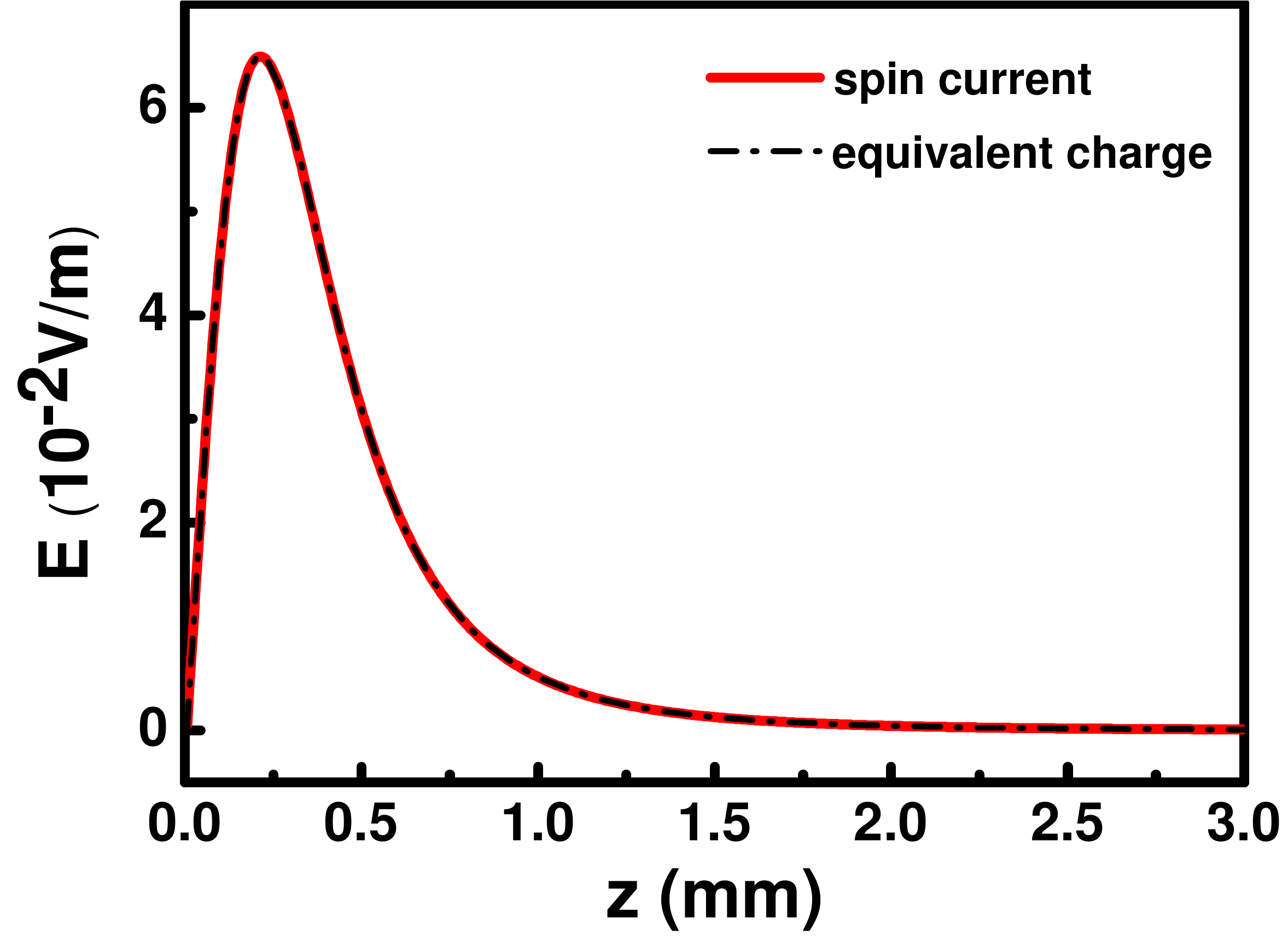}
\caption{(Color online) Electric fields along the positive$\ \rm z$-axis, which are induced by the spin current (red line) and by the equivalent charge (dash dot line).} \label{fig4}
\end{figure}

As the usual spin current, the super spin current can also generate an electric field ${\mathbf E}$ in space.\cite{prb72-245305,prl91-017205,prb69-054409}
Besides, the equivalent charge $\rho$ obtained from the second GL-type equation (\ref{eq:3b}) can produce an additional electric field as well. Therefore, these two electric fields must be identical
if the GL-type theory of the non-polarized spin superconductor is self-consistent. Regarding the polarized spin superconductor, since the relation between the equivalent charge and the super spin current can be explicitly expressed by a formula,\cite{nc4-2951} it can be directly observed that the electric fields calculated from the equivalent charge and from the super spin current are the same. In order to confirm the self-consistency of the non-polarized spin superconductor, we numerically calculate the electric fields. Fig.\ref{fig4} presents the electric fields along the positive$\ \rm z$-axis, which are induced by the equivalent charge (dash dot line) and by the super spin current (red line). As we can see, both electric fields increase first along the positive$\ \rm z$-axis and reach the maximum of $6.5\times10^{-2}\ \rm V/m$ at $0.2\ \rm mm$. Then, they gradually decay to zero. In particular, the perfect superposition between the two curves demonstrates the self-consistency of our theory. Although the electric fields are only presented along the positive$\ \rm z$-axis for clarity, they can be calculated anywhere in space in principle and identical results can be observed. In a word, the electric field induced by the equivalent charge is the same as that generated by the super spin current, which confirms the self-consistency of our theory.

%It is also worthy to note that inside the annulus, the electric field induced by the equivalent
%charge and that induced by the super spin current have opposite direction. This is similar to the case of %a solenoid. Inside a solenoid, the magnetic field induced by charge current is opposite to that induced
%by the equivalent magnetic charge. In that case, charge current is more basic than equivalent magnetic
%charge, and give the right magnetic field inside a solenoid. But in our theory, the equivalent charge is
%more basic than the super spin current according to the second GL-type equation. The equivalent charge, %rather than the spin current, give the right electric field inside the spin superconductor.

\begin{figure}[h]
\includegraphics[width=0.8\columnwidth]{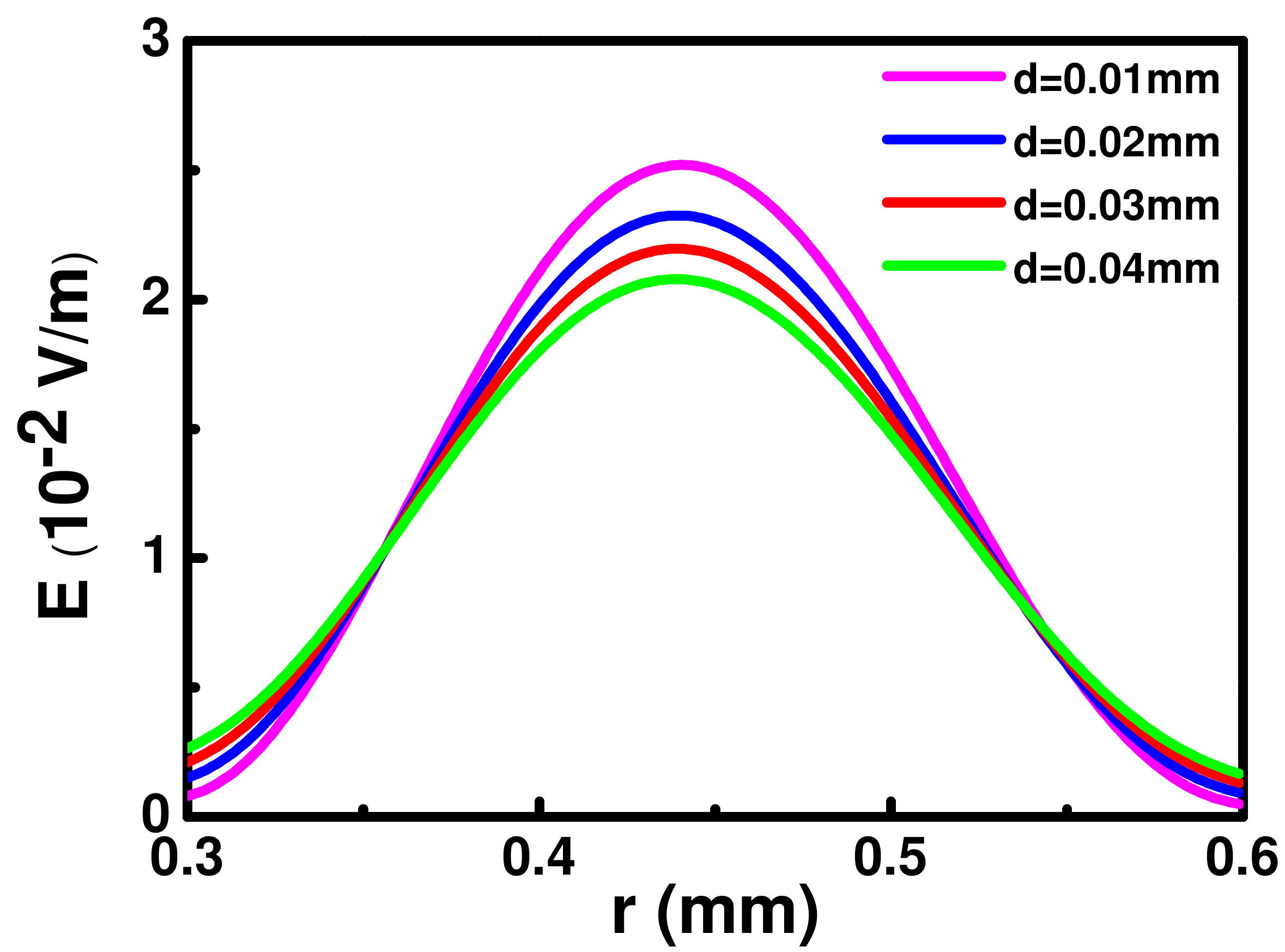}
\caption{(Color online) Induced electric field inside the annulus along the radial direction. Different colors represent different thickness, i.e., d=0.01 mm (pink line), d=0.02 mm (blue line), d=0.03 mm (red line), and d=0.04 mm (green line).} \label{fig5}
\end{figure}

Another interesting issue is the electromagnetic response of the non-polarized spin superconductor under an external electric field. One can see from Fig.\ref{fig2}(a) that the equivalent charge is negative in the upper surface (red line) and positive in the lower surface (blue line). Thus, the equivalent charge in the surface can produce an electric field, of which the direction is opposite to the external electric field which is parallel to the negative$\ \rm z$-axis. Fig.\ref{fig5} shows the numerical results of the electric field induced by the total equivalent charge (both the surface and the bulk). Different colors represent different thickness of the annulus. One can see that when the annulus becomes thicker, the electric field is stronger at the edge but weaker around the middle. It is nearly constant by averaging the induced electric field over the radial direction, irrespective of the thickness of the annulus. This is attributed to the fact that the annulus is a quasi-2D system and the thickness will not contribute to the surface charge density. The magnitude of the induced electric field is approximately $0.02\ \rm V/m$, which is the same order of magnitude as the electric field outside the annulus (Fig.\ref{fig4}). Note that the bulk charge plays an important role on the electric field outside the annulus, whereas the surface charge dominates inside the annulus. This arises from the fact that the surface charge is equivalent to a layer of electric dipoles, where the electric field is strong inside the layer but exhibits a power-law decay outside. In addition, the positive sign of the electric field indicates that the induced electric field tends to screen the external electric field. This is different from the polarized spin superconductor, in which the gradient of the external electric field is screened
and the constant external electric field does not affect it. This difference originates from the spin precession in the non-polarized spin superconductor. Besides, this is also different from metals which can screen the external electric field as well, because the spin superconductor is a charge insulator and no charge current can flow through the system.

\section{\label{sec:conclusions}Summary}

In summary, we generalize the study of spin superconductor to spin-non-polarized case. Since the spin is a vector and can rotate its direction, the order parameter of spin-non-polarized superconductor is a vector in spin space. The free energy of the system is derived and the Ginzburg-Landau-type equations are obtained by using the variational method. When the spin direction is fixed, the Ginzburg-Landau-type equations of the spin-non-polarized superconductor can be reduced to spin-polarized case. In addition, the expressions of super linear and angular spin currents are derived, and the electric field induced by the super spin current is equal to that generated by equivalent charge which is obtained from the second Ginzburg-Landau-type equation, implying the self-consistency of our theory. Finally, by applying these Ginzburg-Landau-type equations, the effect of a constant electric field on the spin superconductor is studied. And the electric fields induced by the super spin current and the equivalent charge are numerically calculated, which can screen the external electric field. These results will help us get a better understanding of the spin superconductor, and the related topics such as Bose-Einstein condensate of magnetic atoms and spin superfluidity.

\section*{Acknowledgement}
We gratefully acknowledge the
support from NSF-China under Grant Nos. 11274364, 11574007, 11504008, and 11504066,
and NBRP of China under Grand Nos. 2012CB921303 and 2015CB921102.
P.L. and Z.Q.B. contributed equally to this work.

\appendix
\section{The derivation of the GL-type equations}
\subsection{\label{sec:eq1a}The first GL-type equation}

In this subsection, we will obtain the first GL-type equation by minimizing the free energy $F_{s}$ with respect to the complex conjugate of the order parameter $\Psi^{\dagger}$.

We first consider the second term $ a\psi_{\mu}^{\ast}\psi_{\mu}$ in Eq.(\ref{eq:2c}). Note that $\int d^3ra\psi_{\mu}^{\ast}\psi_{\mu}=\int d^3ra\Psi^{\dagger}\Psi$,
it can be minimized with respect to $\Psi^{\dagger}$ and we get:
\begin{align}\label{eq:a1}
\delta \int d^3r a\Psi^{\dagger}\Psi =
\int d^3r a\delta\Psi^{\dagger}\Psi .
\end{align}
For the third term in Eq.(\ref{eq:2c}), since $\int d^3r\frac{b_{1}}{2}\psi_{\mu}^{\ast}\psi_{\mu}^{\ast}\psi_{\nu}\psi_{\nu}=\int d^3r\frac{b_{1}}{2}(\Psi^{\dagger}\Psi^{\ast})(\Psi^{T}\Psi)$,
it can be minimized by $\Psi^{\dagger}$ and one gets:
\begin{align}\label{eq:a2}
\delta\int d^3r\frac{b_{1}}{2}(\Psi^{\dagger}\Psi^{\ast})(\Psi^{T}\Psi)
=\int d^3rb_{1}\delta\Psi^{\dagger}\Psi^{\ast}(\Psi^{T}\Psi).
\end{align}
Similarly, by minimizing the fourth term in Eq.(\ref{eq:2c}), i.e., $\frac{b_{2}}{2}\psi_{\mu}^{\ast}\psi_{\nu}^{\ast}\psi_{\mu}\psi_{\nu}=
\frac{b_{2}}{2}(\Psi^{\dagger}\Psi)(\Psi^{\dagger}\Psi)$, one can obtain:
\begin{align}\label{eq:a3}
\delta\int d^3r \frac{b_{2}}{2}(\Psi^{\dagger}\Psi)(\Psi^{\dagger}\Psi)=\int d^3rb_{2}\delta\Psi^{\dagger}\Psi(\Psi^{\dagger}\Psi).
\end{align}
By minimizing the fifth term in Eq.(\ref{eq:2c}), we get:
\begin{align}\label{eq:a4}
&\delta\int d^3r\frac{1}{2m}[({\mathbf p}-\alpha_{0}\mbox{\boldmath $\sigma$}\times{\mathbf E})\Psi]^{\dagger}\cdot({\mathbf p}-\alpha_{0}\mbox{\boldmath $\sigma$}\times{\mathbf E})\Psi \nonumber \\
=&\frac{i\hbar}{2m}\int d^3r\nabla(\delta\Psi^{\dagger})\cdot(-i\hbar\nabla\Psi-\alpha_{0}\mbox{\boldmath $\sigma$}\times{\mathbf E}\Psi) \nonumber \\
&-\frac{\alpha_{0}}{2m}\int d^3r\delta\Psi^{\dagger}(\mbox{\boldmath $\sigma$}\times{\mathbf E})\cdot(-i\hbar\nabla\Psi-\alpha_{0}\mbox{\boldmath $\sigma$}\times{\mathbf E}\Psi).
\end{align}
According to $\nabla\cdot(\varphi{\mathbf f})=(\nabla\varphi)\cdot{\mathbf f} +\varphi\nabla\cdot{\mathbf f}$, we have:
\begin{align}\label{eq:a5}
&\int d^3r\nabla(\delta\Psi^{\dagger})\cdot(-i\hbar\nabla\Psi-\alpha_{0}\mbox{\boldmath $\sigma$}\times{\mathbf E}\Psi) \nonumber \\
=&\int d^3r\nabla\cdot[\delta\Psi^{\dagger}(-i\hbar\nabla\Psi-\alpha_{0}\mbox{\boldmath $\sigma$}\times{\mathbf E}\Psi)] \nonumber \\
&-\int d^3r\delta\Psi^{\dagger}\nabla\cdot(-i\hbar\nabla\Psi-\alpha_{0}\mbox{\boldmath $\sigma$}\times{\mathbf E}\Psi) \nonumber \\
=&\oint d{\mathbf S}\cdot[\delta\Psi^{\dagger}(-i\hbar\nabla\Psi-\alpha_{0}\mbox{\boldmath $\sigma$}\times{\mathbf E}\Psi)] \nonumber \\
&-\int d^3r\delta\Psi^{\dagger}\nabla\cdot(-i\hbar\nabla\Psi-\alpha_{0}\mbox{\boldmath $\sigma$}\times{\mathbf E}\Psi).
\end{align}
We consider the boundary condition:
\begin{align}\label{eq:a6}
[(-i\hbar\nabla-\alpha_{0}\mbox{\boldmath $\sigma$}\times{\mathbf E})\Psi]|_{n}=0,
\end{align}
where the index $n$ means that the direction is perpendicular to
the surface. The boundary condition in Eq.(\ref{eq:a6}) is deduced from the variational principle, which is similar to the
original work by Ginzburg and Landau.\cite{zetf20-1064} Then, we obtain
\begin{align}\label{eq:a7}
&\delta\int d^3r\frac{1}{2m}[({\mathbf p}-\alpha_{0}\mbox{\boldmath $\sigma$}\times{\mathbf E})\Psi]^{\dagger}\cdot({\mathbf p}-\alpha_{0}\mbox{\boldmath $\sigma$}\times{\mathbf E})\Psi \nonumber \\
=&\int d^3r\delta\Psi^{\dagger}\left[-\frac{i\hbar}{2m}\nabla\cdot(-i\hbar\nabla\Psi-\alpha_{0}\mbox{\boldmath $\sigma$}\times{\mathbf E}\Psi)\right. \nonumber \\
&\left.-\frac{\alpha_{0}}{2m}(\mbox{\boldmath $\sigma$}\times{\mathbf E})\cdot(-i\hbar\nabla\Psi-\alpha_{0}\mbox{\boldmath $\sigma$}\times{\mathbf E}\Psi)\right] \nonumber \\
=&\int d^3r\delta\Psi^{\dagger}\frac{1}{2m}(-i\hbar\nabla-\alpha_{0}\mbox{\boldmath $\sigma$}\times{\mathbf E})^2\Psi.
\end{align}
Finally, for the sixth term in Eq.(\ref{eq:2c}),
by minimizing $\int d^3r\Psi^{\dagger}g\mu_{B}\mbox{\boldmath $\sigma$}\cdot{\mathbf B}\Psi$
with respect to $\Psi^{\dagger}$, one gets:
\begin{align}\label{eq:a8}
\delta \int d^3r\Psi^{\dagger}g\mu_{B}\mbox{\boldmath $\sigma$}\cdot{\mathbf B}\Psi =
\int d^3r\delta\Psi^{\dagger}g\mu_{B}\mbox{\boldmath $\sigma$}\cdot{\mathbf B}\Psi.
\end{align}

By combining Eqs.(\ref{eq:a1})-(\ref{eq:a3}), (\ref{eq:a7}), and
(\ref{eq:a8}), we can obtain the first GL-type equation as shown in Eq.(\ref{eq:3a}) in the text.

\subsection{\label{sec:eq1b}The second GL-type equation}

In this subsection, the second GL-type equation is derived by minimizing the free energy $F_{s}$ with respect to the electric potential $\varphi$. It can be seen from Eq.(\ref{eq:2c}) that only the fifth and the seventh terms depend on the electric potential $\varphi$. The derivative of the fifth term over $\varphi$ is:
\begin{align}\label{eq:a10}
&\delta\int d^3r\frac{[({\mathbf p}-\alpha_{0}\mbox{\boldmath $\sigma$}\times{\mathbf E})\Psi]^{\dagger}\cdot({\mathbf p}-\alpha_{0}\mbox{\boldmath $\sigma$}\times{\mathbf E})\Psi}{2m} \nonumber \\
=&\int d^3r\frac{[({\mathbf p}+\alpha_{0}\mbox{\boldmath $\sigma$}\times\nabla\varphi)\Psi]^{\dagger}\cdot(\alpha_{0}\mbox{\boldmath $\sigma$}\times\nabla\delta\varphi\Psi)}{2m} +c.c. \nonumber \\
=&\int d^3r\nabla\delta\varphi\cdot\frac{[({\mathbf p}-\alpha_{0}\mbox{\boldmath $\sigma$}\times{\mathbf E})\Psi]^{\dagger}\times(\alpha_{0}\mbox{\boldmath $\sigma$}\Psi)}{2m}+c.c. \nonumber \\
=-&\int d^3r\delta\varphi\nabla\cdot\frac{[({\mathbf p}-\alpha_{0}\mbox{\boldmath $\sigma$}\times{\mathbf E})\Psi]^{\dagger}\times\alpha_{0}\mbox{\boldmath $\sigma$}\Psi}{2m} +c.c.,
\end{align}
where the boundary condition is taken into account to obtain the last equality.

Similarly, by performing the variational method on the seventh term $\frac{1}{2}\epsilon_{0}{\mathbf E}^2$ in Eq.(\ref{eq:2c}), one obtains:
\begin{align}\label{eq:a11}
&\delta\int d^3r\frac{1}{2}\epsilon_{0}(\nabla\varphi)^2 \nonumber \\
=&\int d^3r\epsilon_{0}\nabla\varphi\cdot\nabla\delta\varphi=\int d^3r\epsilon_{0}[\nabla\cdot(\nabla\varphi\delta\varphi)-\nabla^2\varphi\delta\varphi] \nonumber \\
=&-\int d^3r\delta\varphi\epsilon_{0}\nabla^2\varphi.
\end{align}
Note that the total electric field contains the external electric field and the electric field induced by the super spin current. Since the external electric field is constant as in Ref.[\onlinecite{nc4-2951}], its variation is zero. As a result, one only needs to minimize the free energy with respect to the electric field induced by the super spin current. By combining Eqs.(\ref{eq:a10}) and (\ref{eq:a11}), the second GL-type equation can be obtained straightforwardly (see Eq.[\ref{eq:3b}] in the text).

\end{document}